# Ground Based Support of the Space Mission Parker Performed with Ukrainian Low Frequency Radio Telescopes

Dorovskyy[1], V.V., Melnik[1], V.N., Brazhenko[2], A.I.

[1] Institute of Radio Astronomy of NASU, Kharkiv, 61002, Mystetstv str, 4
[2] Poltava Gravimetric Observatory of Institute of Geophysics of NASU, Poltava, 36014, Myasoyedova str, 27/29.
Email: dorovsky@rian.kharkov.ua, melnik@rian.kharkov.ua, brazhai@gmail.com

## Abstract

Solar sporadic radio emission received in the widest possible frequency band contains important information about the parameters of the sources of this radiation, the solar corona and their variations due to active processes on the Sun. This is what has led to the launches of space missions aimed at studying the Sun and the solar corona, such as Parker Solar Probe (PSP) and Solar Orbiter, in recent years. The purpose of this work is to demonstrate the effectiveness of ground-based support for space missions, primarily PSP, using large Ukrainian decameter radio telescopes. Another goal of the work is to carry out cross-calibration of the radiometers onboard spacecraft using the calibrated data of the ground-based radio telescopes.

One of the most common methods of remote diagnostics of the solar corona is the study of radio emission, the sources of which are located in the solar corona at different heliocentric altitudes. The technique of joint space-terrestrial observations consists in the simultaneous observation of individual events and their analysis in the widest possible frequency band during the maximum approach of the PSP vehicle to the Sun. At the same time, observation in the common frequency band is proposed to be used for calibration of the on-board radio receivers.

The methods of planning joint space-terrestrial observations are substantiated. Using the data of the UTR-2, URAN-2 radio telescopes and the PSP probe, the dynamic and polarization spectra of the simultaneously observed bursts on June 9, 2020 were obtained. The identification and comparison of individual bursts was carried out. A common dynamic spectrum of the bursts in the frequency band 0.5 – 32 MHz was obtained. Cross-calibration of the HFR receiver of the FIELDS-PSP module in the frequency band 10-18 MHz was made using the calibrated data of terrestrial radio telescopes.

The effectiveness of ground-based support of the PSP mission by the large Ukrainian radio telescopes is shown. Examples of joint observations are given, and the method of cross-calibration of the FIELD-PSP module receivers is demonstrated. Prospects for further ground-based support for solar space missions are presented.

**Key words**: radio telescope, UTR-2, space probe PSP, ground-based support, spectral density, flux, calibration.

## Introduction

Ground-based support is an integral part of any space mission, which performs remote sensing of space objects in the radio and optical ranges. This support consists mainly in observing the same events in common wavelength ranges and comparing data obtained by spacecraft with data from ground-based radio telescopes, whose sensitivity, dynamic range, time, frequency, and angular resolutions are apparently much better. Besides purely technical advantages, joint ground-space observations can help in solving many astrophysical problems due to the fact that the simultaneous frequency band in which radio radiation is received may extend from tens of kHz to several GHz, depending on the composition of radio telescopes involved in observations. In particular, when studying the Sun, such ultra-broadband analysis can provide unique information about the properties of the solar corona, sub-relativistic electron beams in an unprecedentedly large range of heliocentric distances - from 1.3 to 50 solar radii almost simultaneously.

Usually, space-borne radio receivers operate in the frequency band from ~10 kHz to 14...16 MHz. From this point of view, technical parameters of the Ukrainian radio telescopes UTR-2, URAN-2 and GURT [1-3] are ideally suited for ground-based support of space missions, as they are able to work in the lowest affordable frequency range for ground observations, limited from below by the ionospheric cutoff frequency (8...10 MHz), and from above with frequencies of 32 MHz (UTR-2, URAN-2) and 70 MHz (GURT), which provides the maximum achievable frequency band available for joint ground-space analysis.

Ukrainian radio telescopes have been actively supporting space missions since 2006, when two identical instruments of the Solar Terrestrial Relations Observatory (STEREO) mission were launched. They were equipped in particular with STEREO-WAVES radio emission recording equipment [4]. Ground-based support for the S-WAVES module of the STEREO mission was planned at the design stage. For this, in addition to the main broad-band sweeping spectrum analysers LFR (Low Frequency Receiver, band 2.5...160 kHz) and HFR (High Frequency Receiver, band 0.125...16.025 MHz), a single-frequency narrow-band receiver FFR (Fixed Frequency Receiver) was also installed. It was tuned to a frequency of 30.025 MHz. The installation of an additional receiver was caused by the lack at that time in Europe and the USA of ground-based instruments capable of reliably receiving radio emissions at frequencies below 30 MHz. This receiver was planned to be used to identify individual bursts observed simultaneously on Earth and in space, with further clarification of the on-board receiving equipment characteristics.

Actually the FFR receiver provided single-frequency profiles of solar radio emissions. Practice shows that due to the morphological diversity of solar radio bursts, their identification only on the base of their profile at a particular frequency is significantly limited. Instead, Ukrainian radio telescopes made it possible to compare the data of the main wide-band onboard HFR receiver with ground-based data at the level of dynamic spectra with significantly better accuracy of identification.

The main directions of support for the STEREO mission were [5]: cross-calibration of space radiometers based on data from Ukrainian radio telescopes in a common frequency band, assistance in the detection of fine-structured bursts, using the ground-based radio telescopes as a third point in the triangulation of solar bursts, clarification of data on the location of radio emission sources received by

STEREO satellites using the gonipolarimetry method [6] with the data of the heliograph based on the UTR-2 [7], the study of the solar wind parameters at large elongations by the IPS method [5]. The wide common frequency band allowed, firstly, to reliably identify an individual burst in the data of different instruments, and secondly, to carry out cross-calibration of the HFR receiver on board the satellite in a wide frequency band, that is very important, bearing in mind the conditions in which space equipment operates. A good example of the expediency of such support is the results of the analysis of the hectometer-decameter type II burst observed by the S-WAVES modules of the STEREO-A/B spacecraft in space and the UTR-2 and URAN-2 radio telescopes on the ground on June 7, 2011 [8].

The main difficulties in the joint processing and analysis of the observational data from ground-based radio telescopes and spacecraft such as WIND and STEREO, are obvious significant difference in the effective area of antennas, and therefore in sensitivity, frequency and time resolutions of the recording equipment.

The time resolution of the spectrum analysers of these satellites is limited by the bandwidth of the data transmission channel from the satellite to Earth and the general limitations of energy consumption. It is usually about 1 minute. Frequency resolution is also significantly limited. This narrowed the range of investigated events mainly by types II and IV bursts, whose duration varies from tens of minutes to several hours. The most common and most numerous type III bursts and their varieties (type IIIb bursts, U-bursts, etc.) were not distinguished in the common frequency band by space spectrum analysers.

The insufficient frequency resolution due to rather slow sweep analysis also limited the list of solar bursts which could be studied. First of all, this applies to fine-structured solar bursts, such as sola spikes, striae (type IIIb bursts), S-bursts, etc. Non-simultaneous, sequential analysis of individual frequencies in the spectrum with a period of about 1 minute significantly distorted the resultant dynamic spectra.

Therefore, even having common frequency bands of ground-based and space-based instruments, the joint data processing was significantly limited.

The situation changed after the launch of NASA's Parker Solar Probe (PSP) on August 12, 2018. The spacecraft was equipped with modern, much more efficient recording equipment. Therefore, ground-based support for these space missions has become even more important.

Chapter 1 of this article describes planning and conducting of joint observational sessions with the Ukrainian radio telescopes and the PSP spacecraft. The details of processing of the data received from the PSP, as well as its comparison with ground-based data are discussed in Chapter 2. The results of the cross-calibration of the space radio telescope based on the observations of a group of powerful bursts of type III, obtained by the UTR-2 radio telescope, are given.

1. **Planning of joint ground-space observations**

Similar to previous solar space missions, such as WIND and STEREO, the PSP probe is aimed at both remote sensing in the radio and optical ranges, as well as in-situ analysis of the parameters of the corona, solar wind, and solar energetic particles (SEP).

In general, there are four separate scientific modules on board the PSP probe, which perform various tasks for the study of the Sun and the near-solar environment [9-12]. In particular, the FIELDS module is designed for the analysis of magnetic and electric fields, temperature and density of the coronal plasma as well as electromagnetic waves in the frequency range of 10.0 kHz...19.2 MHz [9]. From the description given in [9], one can see that the research objects of the FIELDS module of the PSP vehicle and Ukrainian radio telescopes are the same, and their data are obtained in a common frequency range of more than 10 MHz wide.

Unlike previous spacecraft studying the Sun, which were launched into circular orbits either around the Lagrange point (SOHO, WIND) or into the Earth's orbit around the Sun (STEREO) and thus were characterized by a constant distance from the Sun, the PSP satellite has an elliptical orbit with a significant difference between perihelion and aphelion. In addition, its orbit is not constant in time. It gradually shrinks due to periodic flybys of Venus. The perihelion of the orbit decreases from nearly 36 solar radii at the beginning of the mission to less than 10 radii at its end. At the same time, the aphelion decreases from approximately 0.9 to 0.7 astronomical units. The period of rotation of the satellite around the Sun is gradually reduced from 5 months in 2018-2019 to 3 months in 2024-2025.

A characteristic feature of the scientific component of the PSP mission is that the parameters of observations of the Sun change depending on the position of the satellite in orbit with respect to the Sun. In particular, the FIELDS module conducts observations with a maximum time resolution of 3 s only for 12 days around the perihelion time. In the rest of the time (3...5 months), the resolution of the observations equals 56 s, that makes the effective analysis of the ground-space data impossible. In view of the above, ground-based observations by the Ukrainian radio telescopes should only be carried out in an interval of ±6 days from the PSP perihelion date.

In addition, when planning joint observations, it is necessary to take into account the relative positions of the Sun, Earth and PSP during the perihelion, which are different for each perihelion. The major axis of the PSP's orbit ellipse is oriented so that during perihelia in February and March, Earth and PSP are situated on one side from the Sun, while perihelia in late summer are mainly behind-the-limb for an observer on Earth.

The dates of all planned perihelia, the corresponding orbit parameters and the status of ground-based support by the beginning of 2023 are given in Table 1.

The perihelia already provided with ground support by Ukrainian radio telescopes are marked in bold. It should be noted that the PSP probe was launched during the period of the deep minimum of solar activity between the 24th and 25th cycles of activity, in such a way that the lowest perihelion of its orbit will occur in the period of maximum activity of the 25th cycle, which is expected in 2024-25. Because of this, all the first perihelia were characterized by extremely low solar activity with single and mostly weak bursts. This made it difficult to find events suitable for joint analysis.

*Table 1*. Dates of the PSP perihelia and status of the ground-based support by the beginning of 2023.

| No | Perihelion date | Distance from the Sun, Rs | No | Perihelion date | Distance from the Sun, Rs |
|---|---|---|---|---|---|
| 1 | 5 November 2018[2] | 35.6 | 14 | 11 December 2022 | 13.2 |
| 2 | 4 April 2019[2,3] | 35.6 | 15 | 17 March 2023 | 13.2 |
| 3 | 1 September 2019[1,2,3] | 35.6 | 16 | 22 June 2023 | 13.2 |
| 4 | 29 January 2020[1,2,3] | 27.9 | 17 | 27 September 2023 | 11.4 |
| 5 | 7 June 2020[1,2,3] | 27.9 | 18 | 29 December 2023 | 11.4 |
| 6 | 27 September 2020[1,2] | 20,4 | 19 | 30 March 2024 | 11.4 |
| 7 | 17 January 2021[1,2] | 20,4 | 20 | 30 June 2024 | 11.4 |
| 8 | 29 April 2021[1,2,3] | 15.9 | 21 | 30 September 2024 | 11.4 |
| 9 | 9 August 2021[1,2,3] | 15.9 | 22 | 24 December 2024 | 9.9 |
| 10 | 21 November 2021[1,2,3] | 13.2 | 23 | 22 March 2025 | 9.9 |
| 11 | 25 February 2022 | 13.2 | 24 | 19 June 2025 | 9.9 |
| 12 | 1 June 2022[1] | 13.2 | 25 | 15 September, 2025 | 9.9 |
| 13 | 6 September 2022[1] | 13.2 | 26 | 12 December 2025 | 9.9 |

[1] – supported by the URAN-2 radio telescope
[2] – supported by the UTR-2 radio telescope
[3] – supported by the GURT radio telescope

The first cases when an individual solar burst was simultaneously observed by ground-based and space-borne instruments occurred on April 6 and 9, 2019, during perihelion No. 2 and on June 5, 2020, during perihelion No. 5.

A joint analysis of these events based on data from the PSP probe and URAN-2 and GURT radio telescopes is given in [13] and [14], respectively.

## 2. Results of the observations performed within the framework of the ground-based support program

2.1. *Specific features of the data received from FIELDS module of the PSP mission.*

Unlike all previous space radio telescopes, the receiving part of which was based on conventional super-heterodyne receivers with a relatively slow frequency sweep, the registration equipment of the FIELDS module was created using Field-Programmable Gate Arrays (FPGA), which allow performing fast Fourier transform of a broadband signal in real time [9]. This made it possible to expand the working band of radio observations, as well as to significantly increase the resolution, first of all, in terms of time.

The operating bandwidth of the FIELDS frequency analysis is 10.0 kHz...19.2 MHz. At the same time, the overlap ratio of the entire range (the ratio of the maximum frequency of the range to the minimum one) is 1920, which significantly complicates both the acquisition and simultaneous processing of data. Therefore, the range was divided between two receivers - a low-frequency receiver (LFR) operating in a band of 10.0 kHz...1.7 MHz (overlap ratio is 170) and a high-frequency receiver (HFR), which worked in the band of 1.2 MHz...19.2 MHz (overlap ratio is 16). For comparison, the overlap ratios of the UTR-2 and URAN-2 radio telescopes are about 3, and that of the GURT radio telescope equals 7.

In addition, for the highest frequency resolution analysis in the low-frequency range, the LFR receiver of the FIELDS module can operate in a High Resolution mode, in which a 32-point spectrum is obtained in a frequency band of 75 kHz. This provides a frequency resolution of about 2 kHz, that is even higher than the resolution of the DSPZ spectropolarimeter [15] installed at the UTR-2 radio telescope and equals 4 kHz. The central frequency of this high-resolution sub-band is chosen within the whole LFR frequency range by the scientific committee of the mission, depending on the current scientific task.

The original raw data received from the PSP satellite (data level L0) is then processed and transformed into a format convenient for use by the scientific community (levels L1, L2, L3). L2 level data is the closest to the original, which is freely available online. The data of this level is given in the Common Data Format (cdf) developed by NASA. The L2 data appears on the NASA website approximately 6 months after the observation date. The data of the LFR and HFR receivers are stored in separate files.

It should be noted that spectra obtained onboard the PSP satellite have constant relative frequency resolution of about 4-5% across its whole frequency range. Due to this, the frequency channels in the spectrum are placed not in a linear, but in a logarithmic scale. As a result, only 16 of all 64 frequency channels of the HFR receiver fall into the common frequency band with ground-based radio telescopes. This fact causes certain difficulties when comparing PSP and UTR-2 data, which will be discussed further.

### 2.2. *Examples of the solar bursts simultaneously observed by Ukrainian ground-based radio telescopes and the PSP.*

Classic type IIIb-III pairs were registered on April 6 and 9, 2019 by the URAN-2 and GURT radio telescopes and by the PSP spacecraft [13]. According to the URAN-2, the radio emission flux of these bursts at a frequency of 10 MHz were 70 s.f.u. and 350 s.f.u. respectively (1 solar flux unit is equal to 10,000 Jy). Such fluxes do not allow us to classify these bursts as powerful ones [16]. The NOAA 12378 active region, which was most likely responsible for generating the bursts, was located at longitude of 55ºE. At the same time, this active region was almost on the solar limb for the PSP radiometers. Despite the fact that these bursts were reliably recorded by the on-board receiver, the location of the spacecraft with respect to the Earth and the Sun supposes a significant effect of the unknown source radiation pattern on the cross-calibration results. Respectively low flux of these bursts also increased the effect of the galactic background on the cross-calibration results, especially at the lowest frequencies of the common band.

A unique complex event was registered on June 5, 2020 at 9:36 UT by Ukrainian radio telescopes and LFR and HFR receivers of the PSP probe. The event consisted of a group of powerful type III bursts (according to UTR-2, their fluxes reached $10^5$ s.f.u. at frequency of 11 MHz). In the dynamic spectrum these bursts coincide with the ascending branches of a group of solar U-bursts. The results of a joint analysis of this event involving the data from the PSP apparatus and from the GURT radio telescope were considered in [14]. At the time of these observations, the angular separation between the PSP satellite and the Earth with respect to the radio emission source was about 150º. In addition, for the spacecraft, the source was also behind-the-limb [14]. This fact significantly complicates cross-calibration, since the directivity pattern of the source in such a wide angular range is unknown. Moreover, compared to large radio telescopes UTR-2 and URAN-2 the GURT radio telescope conducted observations with only one section of 5x5 dipoles, which yield much smaller effective area and angular selectivity and therefore lower sensitivity and RFI immunity, especially in the common frequency band with the PSP, namely 9...19 MHz. And finally, the GURT data are not calibrated, and can't be used for cross-calibration of the spacecraft.

From the point of view of ground-based support of the PSP space mission, it seems more appropriate to use Ukrainian radio telescopes with large effective area, high angular selectivity, and therefore greater RFI immunity, the data of which are calibrated. First of all, such an instrument is the UTR-2 radio telescope. On the other hand, in order to minimize the effect of the source radiation pattern on the cross-calibration process, it is necessary to select events for which the angle between the source – PSP and source – Earth directions is as small as possible. Obviously, the bursts observed on June 5, 2020 do not meet the above requirements.

The event more closely meeting the requirements of high-quality ground support, was observed on June 9, 2020 at 6:52 UT by the UTR-2 radio telescope. It is a powerful and dense group of type IIIb and III bursts (Fig. 1).

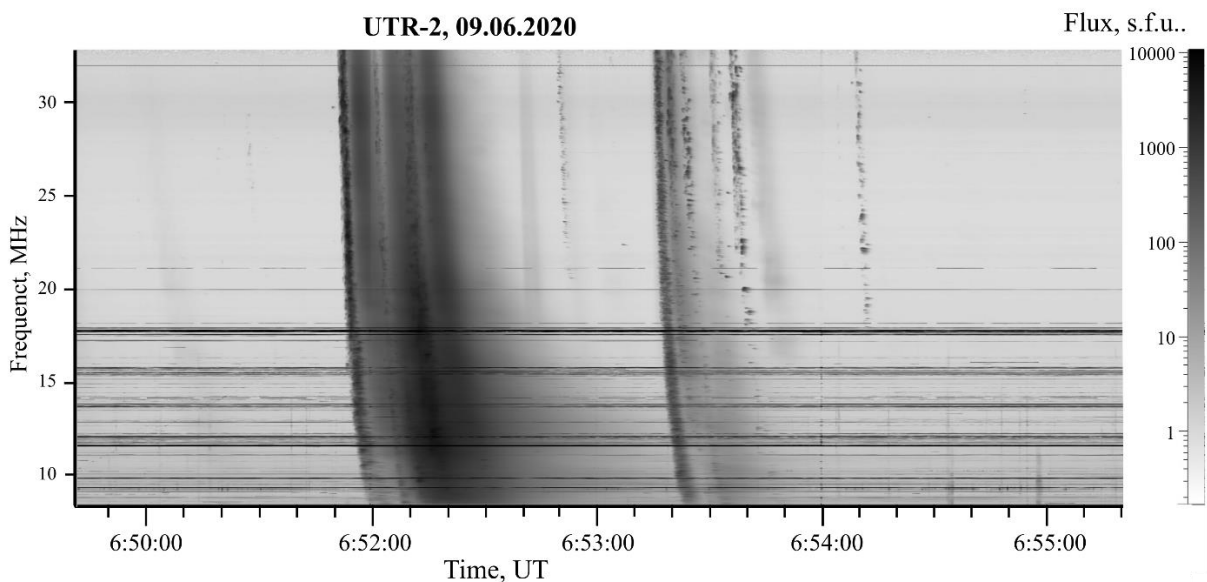

Fig. 1. Group of type IIIb and III bursts (6:52 UT) and group of type IIIb bursts (6:53 UT) observed by UTR-2 radio telescope during the PSP perihelion No. 5.

The flux densities of individual bursts at frequencies below 12 MHz exceed 5000 s.f.u. The duration of this group at a frequency of 30 MHz was about 30 s. This group of bursts was also observed by the FIELDS-HFR receiver on board the PSP at 6:45 UT (frequency 19 MHz, Fig. 2).

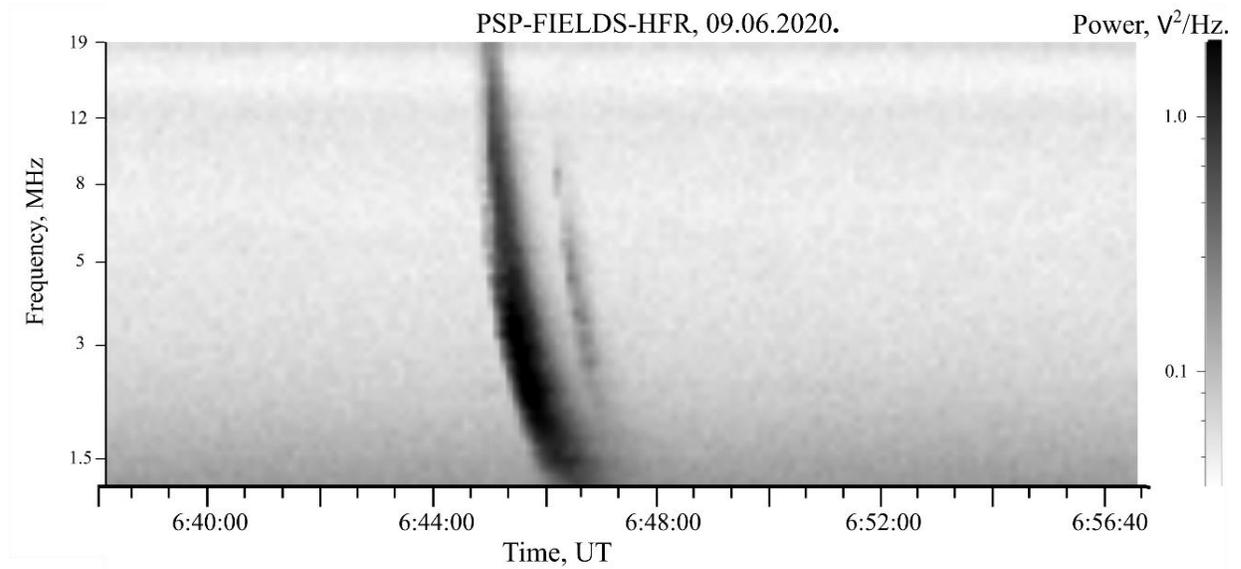

Fig. 2. Group of type III bursts recorded by PSP HFR receiver during the perihelion No. 5.

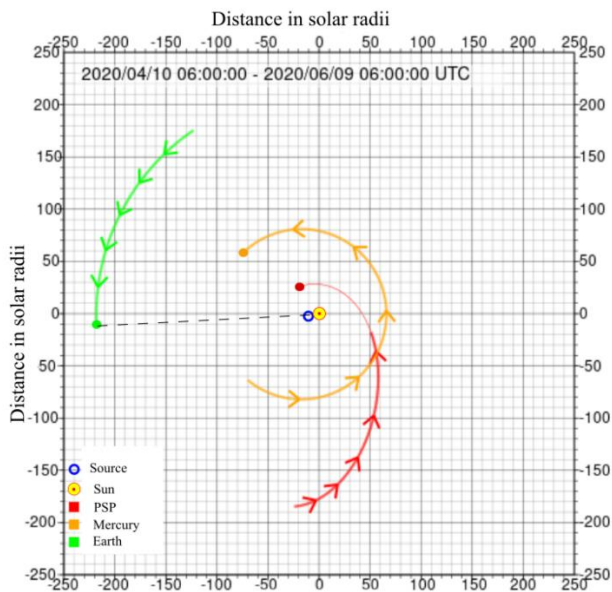
a)

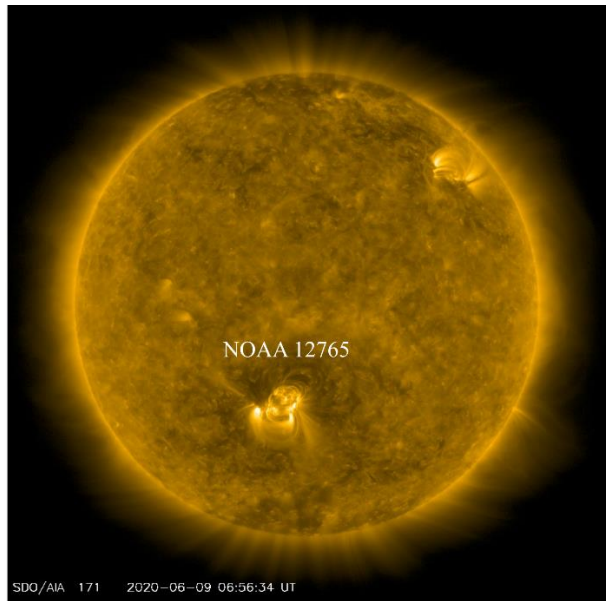
b)

Fig. 3. Relative location of the Earth, PSP spacecraft and the Type III bursts radiation source (a), as well as the position of the active region responsible for the radiation, according to SDO data (b) at the time of observations on June 9, 2020.

Analyzing the images obtained by the Solar Dynamic Observatory (Fig. 3, b), one can assume that the radio emission source of the indicated bursts is associated with the active area of NOAA 12765 located almost at the central meridian. Then the angle between the direction from the source toward the PSP and toward the Earth is about 60º (see Fig. 3, a), which is more acceptable for cross-calibration than in the cases of April 6 and 9, 2019 and June 5, 2020.

Fig. 1 and 2 show that the burst recording times on Earth and onboard the spacecraft are different. This occurred due to the fact that the spacecraft was significantly closer to the source of radio emission than ground-based instruments. Indeed, according to Fig. 3 the heliocentric distance of the PSP was 32 Rs (where Rs is the solar radius), while the Earth was at a distance of 219 Rs. Such a difference in distances corresponds to a delay of the arrival of the signal to the Earth in relation to the PSP of 7 minutes, that is completely consistent with the delay obtained from the dynamic spectra. Joining of UTR-2 and PSP data, taking into account the time delay, resulted in combined dynamic spectrum in the frequency band from 32 MHz to 10 kHz. This spectrum is shown in Fig. 4. Frequencies below 0.5 MHz were removed due to absence of any radio emission at these frequencies.

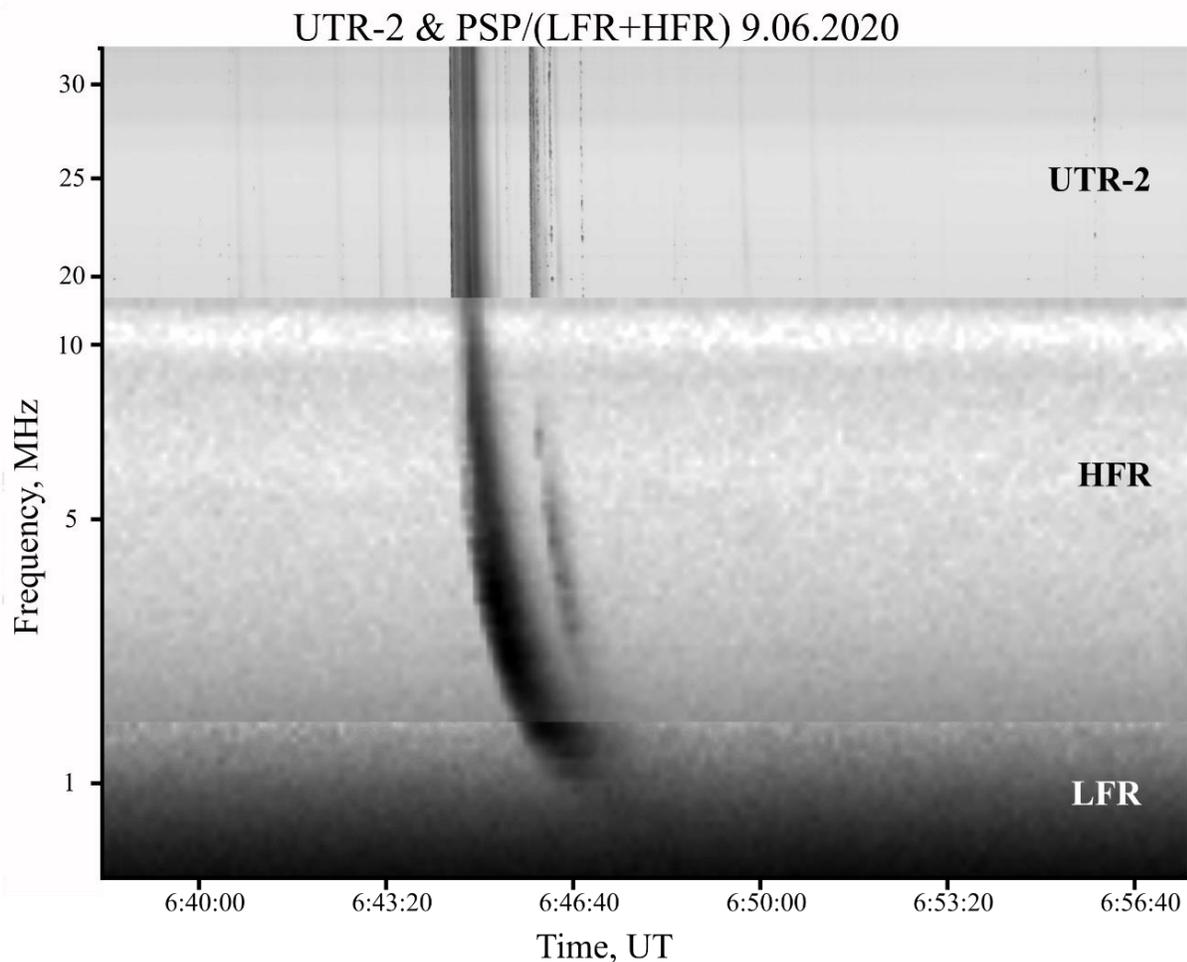

Fig. 4. Combined dynamic spectrum of the analyzed bursts according to the data of the UTR-2 radio telescope and the HFR and LFR receivers of the FIELDS module of the PSP probe.

Fig. 1 shows that approximately one and a half minutes after the group of type III and IIIb bursts, a less powerful group of bursts was observed. The flux densities of these bursts at a frequency of 15 MHz were about 100 s.f.u. The flux density at the location of the PSP probe was estimated to reach 5000 s.f.u. and therefore these bursts had to be reliably recorded by the on-board receiver. However, this group of bursts is completely absent in the PSP data above 9 MHz and is only 4 dB above the Galactic background at 4 MHz.

According to the data obtained with the URAN-2 radio telescope, all type IIIb bursts were strongly polarized, with a degree of circular polarization up to 60%. At the same time, type III bursts are almost unpolarized. This may indicate that the type IIIb bursts emission occurred at the fundamental harmonic of the local plasma frequency, while type III bursts were generated at the second harmonic. Taking into account the fact that the PSP vehicle was located at an angle of about 60º to the likely propagation direction of the source of the analyzed bursts, we can assume that in our case the space probe recorded mainly type III bursts, whose source had a significantly wider radiation pattern than in the case of type IIIb bursts.

### 2.3. *The results of cross-calibration of the PSP-FIELDS-HFR receiver based on the observations of the UTR-2 radio telescope*

Special attention has always been paid to the determination of the spectral flux density of the radio emission received by on-board radiometers of space vehicles, since this parameter is important for finding the mechanism of generation of this emission. The main task of calibrating the space-borne receiving equipment is to retrieve the value of the flux in units of $W \cdot m^{-2} \cdot Hz^{-1}$ from the value of the spectral power density in units of $V^2 \cdot Hz^{-1}$, in which the spectrum is obtained by onboard analyzers. In [17] it was shown that this flux can be calculated as

$$S(f) = C(f) \times P(f) / A_{eff}(f), \qquad (1)$$

where *S(f)* is the spectral flux density at the receiving point, *C(f)* is the factor, *P(f)* is the spectral power density measured by the satellite's on-board spectroanalyzer, *Aeff(f)* is the effective area of the space receiver antenna. All quantities of equation (1) are, in general, frequency-dependent.

The range of wavelengths in which the HFR receiver operates is from 15m to 250m. At the same time, the antenna of the receiver is a dipole with a linear size of 4 m. That is, at all frequencies, the antenna is a short dipole (the length of the dipole is less than half the wavelength). For a short dipole, the effective area does not depend on the linear size of the dipole itself, but is determined exclusively by the electromagnetic wave length [17].

$$A_{eff} = \frac{\lambda^2}{4\pi} \cdot G, \qquad (2)$$

where λ is the wavelength of radio emission, (m), G is the antenna gain, which for a short dipole equals 1.5.

In [17], the galactic background was used to calibrate the data of the RAD2 receiver of the WIND satellite. In this paper we propose to use much more intense sporadic radio emission of the Sun to calibrate the data of the PSP radio receivers. When calibrating with a powerful solar burst signal, the effect of the receiver's own noise and background emission can be neglected.

As mentioned above, a group of type III bursts, observed on June 9, 2020, was selected for cross-calibration of the PSP HFR receiver. In equation (1), the quantity $S(f)$ is the flux density of these bursts at the location of the PSP spacecraft, which has to be found, the quantity $P(f)$ is the spectral power density of the bursts measured at the input of the HFR receiver. It is this value that is contained in the L2 data available online. On the other hand, we have the flux density of these bursts measured by the UTR-2 radio telescope on the Earth's surface ($S_{UTR}$). Since the flux density is inversely proportional to the squared distance from the source, it is possible to obtain the value of the flux density at the location of the satellite, knowing heliocentric distances of the PSP spacecraft and Earth. We can write that

$$S(f) = S_{UTR} \cdot \left(\frac{R_E}{R_{PSP}}\right)^2 = S_{UTR} \cdot K_D^2, \qquad (3)$$

where $R_E$ and $R_{PSP}$ are the heliocentric distances of the Earth and the PSP probe, respectively, and $K_D$ is the ratio of these distances. Combining equations (1-3) we can find the coefficient $C(f)$ as

$$C(f) = \frac{1.5 \cdot S_{UTR} \cdot K_D^2 \cdot \lambda^2}{4\pi \cdot P(f)}. \qquad (4)$$

Thus, knowing the distances of Earth and the space probe from the Sun at the time of observation, the flux density measured by the UTR-2 radio telescope and the spectral power density measured on board the satellite, it is possible to calculate the coefficient $C(f)$ in equation (1) in the entire common frequency band 8...19 MHz.

The values of the coefficient $C(f)$, as well as the corresponding fluxes $S(f)$ and $S_{UTR}(f)$ and the power spectral density $P(f)$ calculated from equation (4) are given in Table 2.

Table 2 shows that the values of the $C(f)$ coefficient at the outermost frequencies of the common band differ significantly (by 1.5...2 times of magnitude) from the rest of the values. For a frequency of 9 MHz, the decrease in the value of the $C(f)$ may be connected with an underestimation of the measured radio emission flux due to absorption in the Earth's ionosphere. The burst was observed 3 hours before the culmination of the Sun. Thus the beam of the radio telescope crossed the ionospheric layer at a significant angle, which increased its thickness.

Frequency of 19 MHz is the upper limit frequency for the HFR receiver, so boundary effects in the antenna matching and filtering circuits of the amplifiers cannot be ruled out. In addition, the effect of the radiation pattern of the source should not be neglected, since the Earth, PSP and the source were still not on the same line.

Table 2. Results of cross-calibration of the FIELDS-HFR receiver of the PSP probe based on the data of observations of the UTR-2 radio telescope.

| Frequency, MHz | $P$, $\times 10^{-15}$ V$^2$·Hz$^{-1}$ | $C$, Ohm$^{-1}$ | $S_{UTR}$, s.f.u. | $S$, s.f.u. |
|---|---|---|---|---|
| 9 | 0.67 | 1.02 | 1117 | 51762 |
| 11 | 0.55 | 1.97 | 2633 | 122000 |
| 13 | 0.46 | 1.62 | 2532 | 117330 |
| 15 | 0.29 | 1.89 | 2485 | 115150 |
| 17 | 0.22 | 1.6 | 2020 | 93790 |
| 19 | 0.18 | 1.24 | 1626 | 75370 |

The data analysis does not show any systematic dependence of the $C(f)$ on frequency. Thus, when the frequency increases by 50%, the variations of the $C(f)$ coefficient amount to only 16% and are rather random in nature. Since the cross-calibration was carried out using the data of only one observation and in a relatively narrow frequency band (11...17 MHz), the authors consider it appropriate to define the estimating coefficient that does not depend on the frequency, having instead a certain accuracy of determination as $C(f)= 1.72\pm0.14$ $\Omega^{-1}$.

As a result, excluding two limit frequencies, we can write

$$S(f) = \frac{(1.72 \pm 0.14) \cdot 4\pi P(f)}{1.5 \cdot \lambda^2}. \tag{5}$$

The coefficient $C(f)$ obtained in this work is definitely an estimate, since it was based on the joint observation of only one burst. Due to the rather narrow common frequency band and the limited number of frequency channels of the FIELDS module in this band, significantly larger observation statistics are required to obtain a reliable dependence of the coefficient $C(f)$ on frequency. Increasing the statistical sample of joint measurements will also contribute to minimizing the effect of the Earth's ionosphere and the radiation pattern of the source on the cross-calibration results. Under conditions of serious damage to the UTR-2 radio telescope, it is quite possible to obtain such statistics during the next PSP perihelia (see Table 1) using the second-largest Ukrainian radio telescope URAN-2.

## Conclusion

The paper considers the issue of ground support for the Parker Solar Probe space mission using large Ukrainian radio telescopes of the decameter range. It is shown that the parameters of UTR-2 and URAN-2 radio telescopes are ideal for joint observations with PSP in the frequency range of 8...19 MHz. The features of space probe observations, the specifics of the data obtained from it, and the methodology of planning joint observations and the features of joint data analysis are substantiated. Examples of simultaneous registration of the same events by ground and space instruments are given. A specific example demonstrates the cross-calibration algorithm of the PSP probe based on data from the UTR-2 radio telescope. Equation (5) is obtained to calculate the flux from the PSP data. Factors that can lead to errors in cross-calibration and methods of their elimination are considered. The expediency of further ground support of the PSP mission by Ukrainian radio telescopes is shown.

**Acknowledgement:** The work was funded within the framework of the state budget grants "Radius" (0122U000616) and "Kronos" (0122U002460). The authors are grateful to the scientific team of the Parker Solar Probe mission for the opportunity to freely use the data obtained during the mission.